\documentclass{article}

\usepackage{PRIMEarxiv}

\usepackage[utf8]{inputenc} 
\usepackage[T1]{fontenc}    
\usepackage{hyperref}       
\usepackage{url}            
\usepackage{booktabs}       
\usepackage{amsfonts}       
\usepackage{nicefrac}       
\usepackage{microtype}      
\usepackage{lipsum}
\usepackage{fancyhdr}       
\usepackage{graphicx}  
\usepackage{amsmath}
\usepackage{subfig}
\graphicspath{{media/}}     

\title{Leveraging Machine Learning for Botnet Attack Detection in Edge-Computing Assisted IoT Networks}

\author{
  Dulana Rupanetti* \& Naima Kaabouch \\
  Artificial Intelligence Research (AIR) Center\\
  University of North Dakota\\
  Grand Forks, ND 58202, USA\\
  \texttt{\{dulana.rupanetti, naima.kaabouch\}@und.edu}
}

\begin{document}
\maketitle

\begin{abstract}
The increase of IoT devices, driven by advancements in hardware technologies, has led to widespread deployment in large-scale networks that process massive amounts of data daily. However, the reliance on Edge Computing to manage these devices has introduced significant security vulnerabilities, as attackers can compromise entire networks by targeting a single IoT device. In light of escalating cybersecurity threats, particularly botnet attacks, this paper investigates the application of machine learning techniques to enhance security in Edge-Computing-Assisted IoT environments. Specifically, it presents a comparative analysis of Random Forest,  XGBoost, and LightGBM —three advanced ensemble learning algorithms—to address the dynamic and complex nature of botnet threats. Utilizing a widely recognized IoT network traffic dataset comprising benign and malicious instances, the models were trained, tested, and evaluated for their accuracy in detecting and classifying botnet activities. Furthermore, the study explores the feasibility of deploying these models in resource-constrained edge and IoT devices, demonstrating their practical applicability in real-world scenarios. The results highlight the potential of machine learning to fortify IoT networks against emerging cybersecurity challenges.
\end{abstract}


\section{Introduction}
\label{introduction}
\begin{figure*}[t]
\centering
\includegraphics[width=13cm]{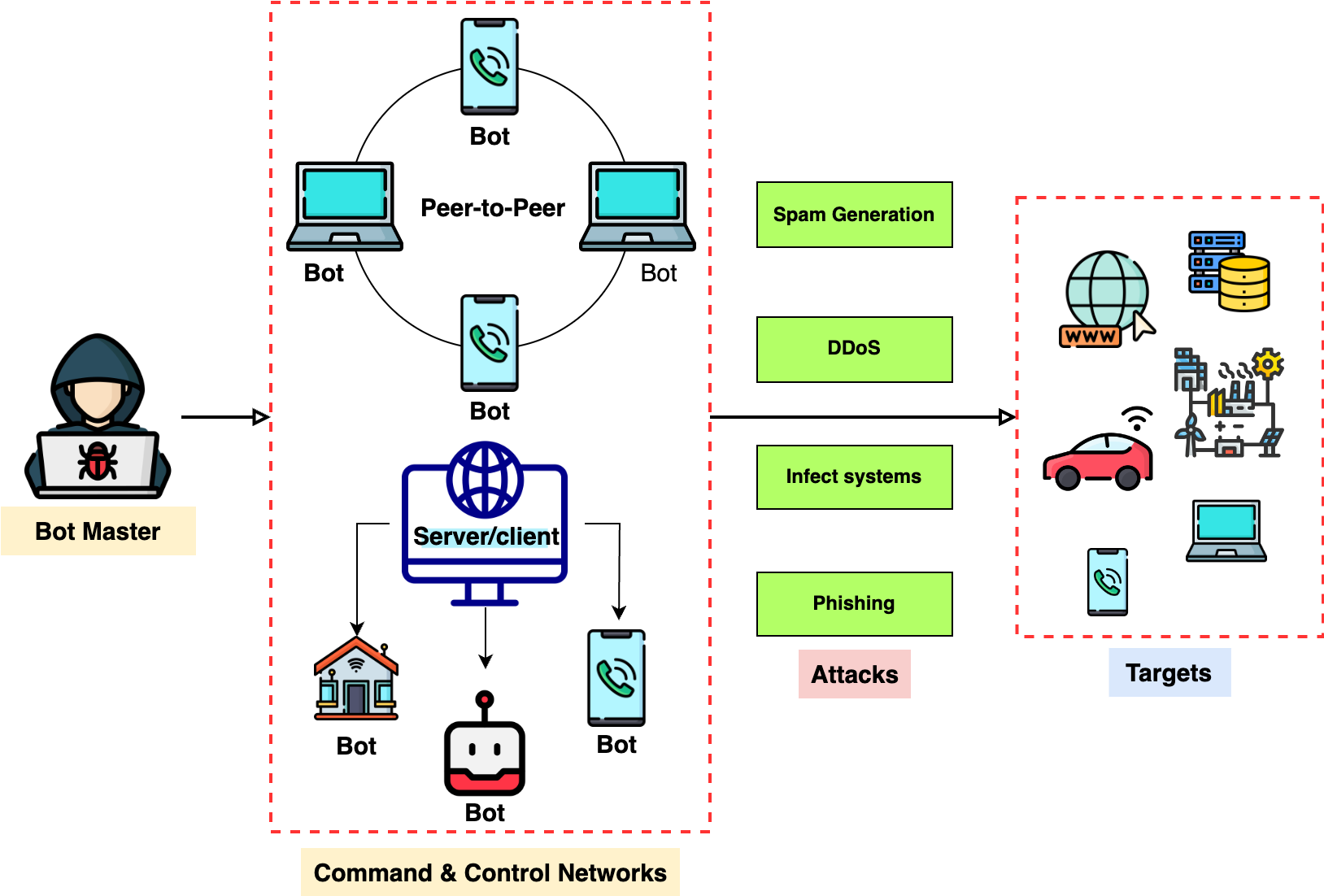}
\caption{Botnet Command \& Control Architecture.}
\label{fig:botmaster}
\end{figure*}

Edge computing has emerged as a transformative paradigm that significantly enhances the functionality and efficiency of Internet of Things (IoT) networks. Processing data closer to the source—at the edge of the network—addresses critical challenges associated with traditional cloud computing, such as latency, bandwidth limitations, and security concerns. The integration of edge computing into IoT networks allows for real-time data processing and decision-making, which is essential for applications that require immediate responses, such as smart cities, healthcare, and autonomous vehicles \cite{salh2023energy, bukhsh2021decentralized, chen2019iot}.

One of the primary advantages of edge computing in IoT networks is the reduction of latency. By enabling data processing at or near the source, edge computing minimizes the time it takes for data to travel to centralized cloud servers and back. This is particularly important for time-sensitive applications, where delays can lead to significant consequences. For instance, in healthcare applications, edge computing can facilitate real-time monitoring and analysis of patient data, thereby improving response times in critical situations \cite{salh2023energy, bukhsh2021decentralized, phuc2022traffic}.

Furthermore, edge computing eases the burden on cloud infrastructure by offloading processing tasks to local edge nodes, which can handle computations more efficiently and reduce the overall network traffic \cite{jin2020secure, jiang2019toward}. Security is another critical aspect where edge computing plays a vital role. By processing sensitive data locally, edge computing reduces the risk of data exposure during transmission to the cloud. This is especially relevant in IoT environments where numerous devices generate vast amounts of data that may contain personal or sensitive information. Implementing security measures at the edge can enhance data protection and privacy, allowing for more localized control over data management \cite{xu2021enhanced, hsu2020privacy, pathak2023tabi}. Furthermore, decentralized resource allocation strategies in edge computing can further strengthen security by minimizing the risks associated with centralized data storage and processing \cite{sasikumar2023decentralized}.

Edge computing supports IoT network scalability by enabling the integration of various devices and applications. This is particularly beneficial in heterogeneous IoT environments, where devices with contrasting capabilities and requirements coexist. Edge computing facilitates seamless communication and coordination among these devices, enhancing the overall system's performance and reliability \cite{lan2019iot, wang2019enhancing}. The ability to dynamically allocate resources at the edge also ensures IoT applications can adapt to changing demands and conditions, improving their efficiency and effectiveness \cite{kim2020collaborative}.

The upsurge of IoT devices has significantly increased the attack surface for cybercriminals, mainly through exploiting botnets. Botnet attacks on IoT networks use compromised devices that attackers remotely control to execute various malicious activities. These attacks have evolved in complexity and scale, posing serious threats to individual users and broader network infrastructures. One of the most notable types of botnet attacks is the Distributed Denial of Service (DDoS) attack, where a network of compromised IoT devices overwhelms a target server with traffic, rendering it unavailable to legitimate users. Reports indicate that botnets like Mirai and BASHLITE have been instrumental in executing large-scale DDoS attacks, with the number of IoT devices involved in such attacks rising dramatically in recent years. For instance, a recent threat intelligence report noted an increase from approximately 200,000 to 1 million IoT devices participating in botnet-driven DDoS attacks within a year \cite{selvam2024uasdac}.

This surge is primarily attributed to the inherent vulnerabilities of IoT devices, many of which lack adequate security measures, making them easy targets for exploitation \cite{alshehri2024skipgatenet, hussain2021twofold}. Furthermore, the methods employed by botnets have diversified. Attackers can utilize compromised IoT devices for various malicious activities, including credential stuffing, identity theft, and spamming \cite{panda2021efficient}. The ability of botnets to execute simultaneous attacks across numerous devices amplifies their impact, complicating detection and mitigation efforts. These botnets' command-and-control infrastructure, as shown in Figure \ref{fig:botmaster}, allows attackers to issue commands to infected devices, facilitating data theft and manipulation \cite{sattari2022hybrid, taher2023reliable}. This decentralized control model makes it challenging for security systems to identify and neutralize threats effectively.

The rapid evolution of attack strategies further complicates the detection of botnet attacks in IoT environments. As attackers continuously adapt their methods, traditional security measures often fall short. For instance, the emergence of low-rate DDoS attacks, which aim to evade detection by generating minimal traffic, poses a significant challenge for existing detection systems \cite{liu2020cpss}. Moreover, the diversity of IoT devices with unique hardware and software configurations necessitates tailored detection strategies to identify compromised devices effectively \cite{trajanovski2021framework, stephens2021detecting}.

Researchers are exploring advanced detection techniques leveraging machine learning (ML) and deep learning (DL) to combat botnet threats by analyzing network traffic patterns and identifying anomalies indicative of malicious activity. Hybrid models that combine various ML techniques have shown promise in achieving high detection accuracy \cite{ali2024hybrid}. However, while DL approaches are exceptionally accurate, their substantial computational and memory requirements, often necessitating specialized hardware such as GPUs, make them less suitable for resource-constrained environments like IoT-edge networks. In such scenarios, lightweight ensemble-based ML models, such as Random Forest, XGBoost, and LightGBM, offer a practical alternative by balancing detection accuracy with computational efficiency, enabling real-time monitoring and response through resource-efficient Intrusion Detection Systems \cite{arnold2024network}.

This work focuses on botnet detection at the edge level using machine learning models optimized for resource-constrained environments. Three models—Random Forest, XGBoost, and LightGBM—were implemented and evaluated using a widely recognized IoT botnet dataset. The models were trained and tested on both preprocessed and noisy data to assess their performance and suitability for deployment on edge or IoT devices. Additionally, a Deep Feedforward Neural Network (DFNN) was implemented to compare its effectiveness against the ML models, providing insights into the impact of DL architectures in IoT edge environments.

The remainder of this paper is structured as follows: Section \ref{relatedwork} provides the current work in the literature related to botnet attacks on the edge. Section \ref{methods} discusses the proposed methodology. Section \ref{experiment} describes the experiment and the results. Finally, section \ref{conclusion} contains the conclusion and future directions.

\section{Related Work}
\label{relatedwork}
Botnets continue to pose a critical threat to edge networks, prompting extensive research into efficient detection and mitigation methods rooted in ML. A key motivation behind these studies is addressing the resource limitations inherent to edge and IoT devices, which often challenge computationally heavy DL techniques.

Several early works have highlighted the value of ML in botnet detection and outlined its fundamental role in analyzing network traffic. For instance, Stevanovic and Pedersen \cite{stevanovic2016use} provided a comprehensive survey of ML-based botnet detection methods, underscoring how ML can effectively capture subtle behaviors that distinguish botnets from regular traffic. Yang et al. \cite{Yang2019} went on to propose a feature extraction method leveraging the concept of graphic Symmetry for detecting peer-to-peer (P2P) botnets, even when communications are encrypted—an issue that remains central in modern detection strategies.

Building on the potential of ML, many researchers explored advanced ML and  DL techniques and privacy-aware frameworks to safeguard IoT-edge environments better. In particular, Popoola et al. \cite{Popoola2022} developed a Federated DL (FDL) model that addresses zero-day botnet attacks on IoT-edge devices while mitigating data privacy leakage through decentralized training. Similarly, Salman et al. \cite{Salman2019} proposed an ML-based framework for identifying IoT devices and detecting abnormal traffic in edge networks, emphasizing the importance of ensuring both security and privacy. Complementing these efforts, Sedjelmaci et al. \cite{Sedjelmaci2022} introduced a trusted hybrid learning model that fuses human security experts’ domain knowledge with ML algorithms to defend against both known and emerging threats in edge networks.

Other studies have focused on direct mitigation strategies, highlighting how ML and DL can be pivotal in stopping botnet attacks before they inflict widespread damage. For example, Yerima et al. \cite{Yerima2022} demonstrated a DL-enhanced system that relies on text mining Android manifest files to detect Android-specific botnets. Chu et al. \cite{Chu2019} presented a sophisticated algorithm combining network vulnerability analysis with ML to refine evaluation and screening processes, thereby bolstering botnet detection mechanisms. In a related vein, Shareef et al. \cite{Shareef2024} harnessed advanced DL techniques, such as Dual-channel Graph Attention Networks and the Zebra Optimization Algorithm, to enhance detection accuracy. While these methods achieved strong performance, issues of scalability, implementation complexity, and model generalizability often limit their practical deployment on resource-constrained edge devices.

Addressing this same limitation, Alissa et al. \cite{Alissa2022} proposed a combined feature extraction and classification ML methodology for botnet detection in IoT networks, which relies heavily on packet-level analysis. Although effective, its reliance on real-time packet inspection and challenges in handling encrypted traffic restrict its deployment in specific modern IoT ecosystems. Similarly, Alqahtani et al. \cite{Alqahtani2022} introduced an ML-based framework called EDIMA for early detection of IoT malware activities. By analyzing network traffic patterns during the crucial scanning and infecting phases of botnet formation, EDIMA can proactively prevent IoT devices from being compromised. Nonetheless, real-time detection and encrypted traffic remain open hurdles.

Doshi et al. \cite{Doshi2018} studied botnet-driven DDoS attacks emanating from consumer IoT devices to enhance network-level detection. Their ML framework—which leverages Random Forest and Neural Networks—analyzes features like packet size and inter-packet intervals, showing high accuracy for identifying botnet traffic. Likewise, Mahajan et al. \cite{Mahajan2023} developed an unsupervised DL model using autoencoders to learn normal device behavior. Reconstruction errors signaled potential botnet activities, displaying the capacity to outperform traditional rule-based techniques. In another recent contribution, Rabhi et al. \cite{Rabhi2023} tackled Mirai botnet detection using DL and mining techniques; however, the model’s reliance on network traffic data poses challenges in managing encrypted communication. Finally, Bakhshad et al. \cite{Bakhshad2022} combined feature selection using the NSL-KDD dataset with deep reinforcement learning, demonstrating how hyperparameter optimization can boost detection accuracy for botnet mitigation.

Despite the increase in ML and DL solutions, their practicality on edge devices remains questionable due to the high computational and storage demands. Edge and IoT devices often have minimal memory and processing capacity, making the deployment of large-scale DL models highly impractical. Against this backdrop, this paper introduces an alternative using decision tree-based algorithms -- Random Forest,  XGBoost, and LightGBM, comparatively lightweight and interpretable ML algorithms. Their hierarchical structure eases computational overhead and enables flexible adaptation to diverse hardware architectures, offering a compelling solution for robust botnet detection in resource-constrained edge environments.

\section{Methodology}
\label{methods}
\begin{table}[]
    \centering
    \begin{tabular}{lp{10cm}}
        \toprule
        \textbf{Feature} & \textbf{Description} \\
        \midrule
        \texttt{id.orig\_p} & Source port used by the originator of the connection. \\
        \texttt{id.resp\_p} & Destination port used by the responder in the connection. \\
        \texttt{proto} & Network protocol used (e.g., TCP, UDP, ICMP). \\
        \texttt{service} & Application-layer service (e.g., HTTP, DNS). \\
        \texttt{duration} & Duration of the connection in seconds. \\
        \texttt{orig\_bytes} & Total bytes sent by the originator. \\
        \texttt{resp\_bytes} & Total bytes sent by the responder. \\
        \texttt{conn\_state} & State of the connection (e.g., established, reset). \\
        \texttt{history} & State history of the connection (e.g., flags for data transfer, resets). \\
        \texttt{orig\_pkts} & Number of packets sent by the originator. \\
        \texttt{orig\_ip\_bytes} & Total IP-layer bytes sent by the originator. \\
        \texttt{resp\_pkts} & Number of packets sent by the responder. \\
        \texttt{resp\_ip\_bytes} & Total IP-layer bytes sent by the responder. \\
        \texttt{label} & classification of the connection (e.g., Attack, Benign). \\
        \bottomrule
    \end{tabular}
    
        \caption{Explanation of dataset features.}
        \label{tab:datasetdescription}
\end{table}

This section provides a detailed discussion of the attack scenario, the dataset, the data preprocessing techniques employed, the classifiers used, and the performance metrics. Additionally, it covers the feature selection process and the hyperparameter tuning used for optimal model training and evaluation.

\subsection{Dataset}

Using a clean and reliable dataset is essential to developing an accurate and well-trained ML model. This study focuses on a subset of the IoT-23 dataset, as detailed in \cite{garcia2020iot23}. The IoT-23 dataset, released in January 2020, contains network traffic data from IoT devices, consisting of 20 malicious activities and three instances of benign traffic. The data was collected between 2018 and 2019 by the Stratosphere Laboratory at the Artificial Intelligence Center (AIC), Faculty of Electrical Engineering (FEL) at CTU University in Prague, Czech Republic, with support from Avast Software.

In the IoT-23 dataset, each sample includes details on capture duration, packet count, Number of Zeek flows, and data size, representing different malware types such as Mirai, Torii, Trojan, Gagfyt, Okiru, Kenjiro, Hakai, IRCBot, Muhstik, and Hide and Seek. The table \ref{tab:datasetdescription} better explains each feature of a sample in this dataset. The dataset is extensive and contains over 6 million samples. This work only looks at 400,000 samples, split between training, evaluation, and testing.

\begin{figure}
\centering
\includegraphics[width=10cm]{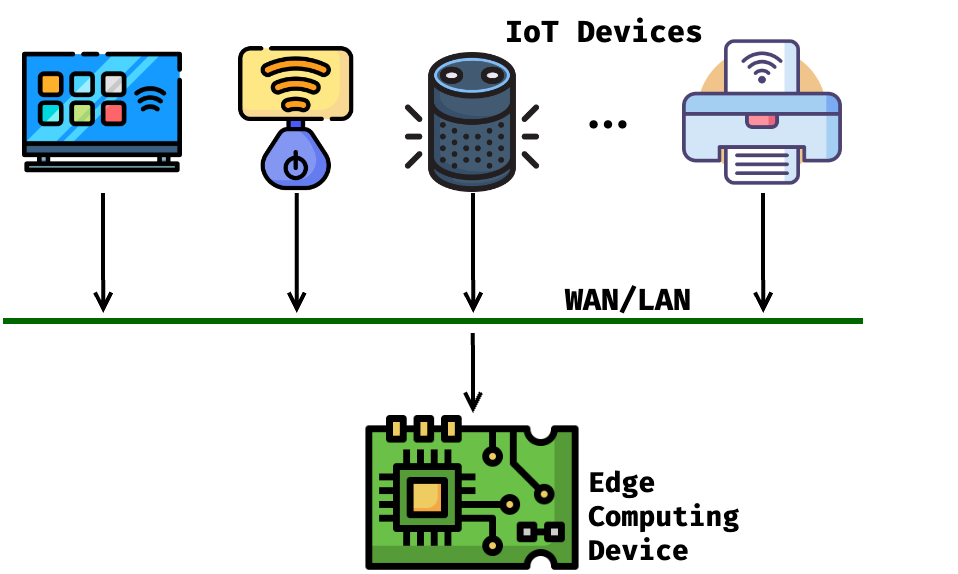}
\caption{High-level view of IoT Devices connected to an Edge Device.}
\label{fig:device-connect}
\end{figure}

A high-level schematic of IoT devices connected to an edge device is shown in Figure~\ref{fig:device-connect}.

\subsection{Attack Scenarios}

The attack scenarios in the selected IoT-23 dataset involve executing malware on IoT devices such as Raspberry Pi and capturing traffic from malware like Mirai, Torii, Hide and Seek, Hajime, and others. The dataset provides labeled network traffic, including detailed characteristics and labeling based on human analysis. It includes both .pcap files and connection logs for analysis, which helps develop security algorithms for IoT devices.

\subsection{Data pre-processing}

\begin{figure}[!b]
\centering
\includegraphics[width=8.5cm]{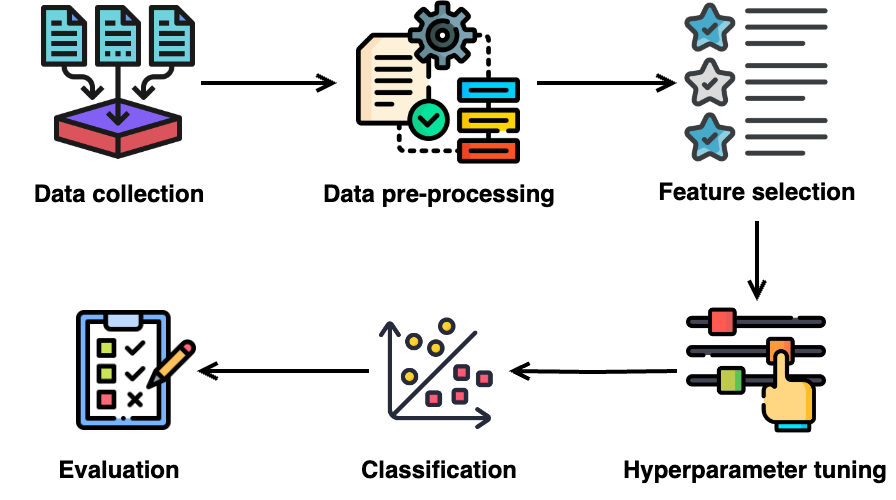}
\caption{Botnet Detection: Data Collection, Feature Extraction, Classification, and Tuning.}
\label{fig:}
\end{figure}

Data preprocessing is a fundamental step in the ML pipeline, ensuring that raw data is transformed into a clean, structured format suitable for modeling. This process is essential for improving data quality, which impacts the performance and reliability of ML algorithms. In the context of this study, preprocessing ensures that the dataset is optimized for analysis and prediction, especially when dealing with large and diverse data repositories.

The primary goals of data preprocessing include cleaning the dataset by removing noise, resolving inconsistencies, and handling missing values. These steps enhance the dataset's integrity, making it suitable for downstream modeling. Additionally, feature extraction and selection are employed to identify and isolate the most informative variables, reducing dimensionality and ensuring that the model focuses on relevant data. Normalization and scaling are also applied to standardize features, further improving the model's ability to generalize across varied data distributions, particularly in resource-constrained environments like IoT-edge devices.

\subsection{Feature Selection}

Feature extraction is a critical step in ML that significantly influences the performance and efficiency of models. Transforming raw data into a more manageable set of relevant features helps reduce dimensionality, simplifying analysis and visualization while retaining essential information. It also helps to identify any issues in processing a more manageable data set with relatively clean data. Moreover, effective feature extraction can improve model accuracy and generalization by focusing on the most informative aspects of the data, thereby mitigating the risk of overfitting. 

In this work, we have performed feature selection using Spearman's rank correlation and feature importance using XGBoost methods.

\subsubsection{Spearman's rank correlation}

Spearman's rank correlation coefficient (SRCC) is a non-parametric measure that assesses the strength and direction of association between two ranked variables. Spearman's correlation evaluates the relationship based on the ranks of the data rather than their actual values. This characteristic makes it particularly useful in scenarios where the data may not meet the assumptions necessary for parametric tests, such as in the presence of outliers or non-linear relationships \cite{liu2022analysis, li2020interactive}. The SRCC is calculated by determining the rank difference for each pair of observations, allowing for a more robust analysis of the correlation between variables. Spearman's rank correlation in ML modeling can significantly enhance feature selection and model performance. The equation \ref{eq:spearman} defines the Spearman
rank correlation coefficient where $X_r$ and $Y_r$ represent the ranked values of the random variables $X$ and $Y$, respectively; $Cov()$ denotes the covariance, and $Std()$ refers to the standard deviation.
 
\begin{equation}
   r = \frac{\mathrm{Cov}(X_r, Y_r)}{\mathrm{Std}(X_r) \cdot \mathrm{Std}(Y_r)} 
   \label{eq:spearman}
\end{equation}

\subsubsection{XGBoost Feature Importance}
In XGBoost, feature importance measures the contribution of each feature to the model's predictions. It helps identify which features significantly impact model accuracy by evaluating their role in splitting the decision trees that XGBoost constructs. XGBoost provides several methods to assess feature importance, such as "gain," which measures the improvement in accuracy brought by a feature; "cover," which reflects the Number of observations affected by a feature; and "weight," which counts how often a feature is used in splits.

\subsection{Machine Learning Models}

This paper focuses on implementing an ML solution on the edge of an IoT network to detect botnet attacks. Therefore, the chosen ML model has to be lightweight and less resource-hungry while maintaining an acceptable level of accuracy in detecting botnets. Considering the requirements, we decided on Random Forest Classifier, XGBoost, and LightGBM to implement the ML models and compare the performance metrics.

Table \ref{tab:model_comparison} describes a high-level comparison of these models. The table compares Random Forest, XGBoost, and LightGBM across accuracy, speed, overfitting risk, ease of use, and scalability. Random Forest is easy to use but less scalable, XGBoost offers high accuracy at a computational cost, and LightGBM excels in speed and scalability but requires careful tuning. The following sub-sections discuss these models in detail.

\begin{table}[]
\centering
\begin{tabular}{@{}lccc@{}}
\toprule
\textbf{Feature}                  & \textbf{Random Forest }      & \textbf{XGBoost}            & \textbf{LightGBM}           \\ \midrule
\textbf{Accuracy}        & Moderate to Good    & High               & Very High          \\
\textbf{Training Speed}  & Moderate            & Slow               & Fast               \\
\textbf{Prediction Speed}& Moderate            & Moderate           & Fast               \\
\textbf{Overfitting Risk}& Low                 & Moderate (with tuning) & Moderate to High  \\
\textbf{Ease of Use}     & Easy & Moderate  & Moderate         \\
\textbf{Scalability}     & Moderate            & High               & Very High          \\ \bottomrule
\end{tabular}
\caption{High-level comparison of Random Forest, XGBoost, and LightGBM models.}
\label{tab:model_comparison}
\end{table}

\subsubsection{Random Forest}

Random Forest is an ML method that forges an ensemble of decision trees, each trained on a randomly chosen subset of the data and features. For each prediction, it combines the outputs of all trees: using the majority vote in classification tasks and the average value for regression. This technique is known for its robustness to overfitting and its ability to handle datasets with many features. To construct each tree, data samples are drawn with replacement (bootstrap sampling), and at every split, only a random selection of the total features is considered. Typically, for classification problems, the number of features examined at each split is set to \(\sqrt{p}\), where \(p\) represents the total feature count.

The forest aggregates predictions from individual trees. For classification, the final prediction is based on majority voting:

\begin{equation}
    \hat{y} = \mathrm{mode}(\hat{y}_1, \hat{y}_2, \ldots, \hat{y}_M)
\end{equation}

where $\hat{y}_j$ represents the prediction of the $j$-th tree, and $M$ is the total number of trees. For regression, the prediction is the mean of individual tree outputs:

\begin{equation}
\hat{y} = \frac{1}{M} \sum_{j=1}^M T_j(x)
\end{equation}

where $T_j(x)$ is the prediction of the $j$-th tree for input $x$.

During training, the algorithm determines splits using a criterion such as Gini Impurity, which measures the homogeneity of data within a node. For a node $S$, Gini Impurity is defined as:

\begin{equation}
\text{Gini}(S) = 1 - \sum_{k=1}^K p_k^2
\end{equation}

where $p_k$ is the proportion of samples in $S$ belonging to class $k$, and $K$ is the number of classes. The algorithm selects splits that minimize the weighted Gini Impurity of child nodes, calculated as:

\begin{equation}
\Delta \text{Gini} = \text{Gini}(S) - \frac{|S_L|}{|S|} \text{Gini}(S_L) - \frac{|S_R|}{|S|} \text{Gini}(S_R)
\end{equation}

where $S_L$ and $S_R$ are the left and right child nodes, and $|S|$ is the number of samples in $S$. This process iteratively refines splits to optimize predictive performance.

\subsubsection{XGBoost}

XGBoost, introduced by Chen and Guestrin~\cite{Chen2016}, is an efficient and scalable implementation of Gradient Boosting Decision Trees (GBDT). It constructs an ensemble of weak learners (usually decision trees), where each tree is trained to correct the errors of its predecessors. This iterative process yields a high-performing model favored in many ML competitions and industrial applications.

The key to XGBoost's success lies in its objective function, which combines a loss term \(l\) (measuring predictive error) with a regularization term \(\Omega\) (penalizing model complexity):
\begin{equation}
\label{eq:obj_func}
\text{Obj} 
= \sum_{i=1}^n l\bigl(y_i, \hat{y}_i\bigr) 
\;+\; \sum_{m=1}^M \Omega\bigl(f_m\bigr),
\end{equation}
where \(l(\cdot)\) can be any differentiable convex loss (e.g., mean squared error or cross-entropy), and \(\Omega(f_m)\) mitigates overfitting by penalizing overly complex trees.

At each boosting iteration \(t\), XGBoost updates its predictions by adding a new tree \(f_t\):
\begin{equation}
\label{eq:incremental_update}
\hat{y}_i^{(t)} = \hat{y}_i^{(t-1)} + f_t(x_i),
\end{equation}
Where \(f_t\) is learned based on first- and second-order gradients of the loss with respect to \(\hat{y}_i\). This gradient-based approach utilizes more detailed information than standard boosting, accelerating convergence and enhancing accuracy.

The final prediction is the sum of the contributions from all \(M\) trees:
\begin{equation}
\label{eq:final_pred}
\hat{y}_i = \sum_{m=1}^M f_m(x_i).
\end{equation}

In addition to its gradient-based tree-building strategy, XGBoost incorporates several optimizations such as parallel tree construction, efficient handling of missing values, and a quantile sketch for weighted data. These enhancements allow XGBoost to scale effectively to large, sparse, and high-dimensional datasets.

\subsubsection{LightGBM}

LightGBM, introduced by Ke \emph{et al.}~\cite{Ke2017}, is a gradient-boosting framework designed with a primary focus on efficiency and scalability. It builds an ensemble of decision trees iteratively to minimize predictive error and is highly regarded for its speed, memory efficiency, and ease of handling categorical features. These attributes make LightGBM particularly well-suited for large datasets with high-dimensional feature spaces.

Similar to other gradient boosting methods, the LightGBM objective function includes both a loss term \(\,l\), measuring prediction error, and a regularization term \(\,\Omega\), penalizing model complexity to prevent overfitting:
\begin{equation}
\label{eq:obj_func_lgbm}
    \text{Obj} 
    = \sum_{i=1}^n l\bigl(y_i, \hat{y}_i\bigr) \;+\; \sum_{m=1}^M \Omega\bigl(f_m\bigr).
\end{equation}

Several key optimizations distinguish LightGBM from other boosting frameworks:
\begin{itemize}
    \item \emph{Gradient-Based One-Side Sampling} (GOSS) reduces training set size by prioritizing data points with large gradients.
    \item \emph{Exclusive Feature Bundling} (EFB) combines mutually exclusive features into a single feature, lowering dimensionality without sacrificing information.
    \item A \emph{histogram-based} training algorithm accelerates feature discretization and reduces memory usage.
\end{itemize}

Once the ensemble of \(M\) trees is trained, the final prediction for each sample \(x_i\) is the sum of all tree outputs:
\begin{equation}
\label{eq:final_pred_lgbm}
    \hat{y}_i = \sum_{m=1}^M f_m(x_i).
\end{equation}

These design choices enable LightGBM to handle high-volume, high-dimensional data while maintaining competitive accuracy and speed relative to other gradient-boosting implementations.

\subsection{Peroformance metrics}

Precision, recall, and F1-score are essential metrics for evaluating the performance of ML models, particularly in scenarios with imbalanced datasets. Precision quantifies the accuracy of positive predictions, while recall measures the model's ability to identify all relevant instances. The F1-score, the harmonic mean of precision and recall, provides a single metric balancing both aspects. This makes it especially valuable when dealing with skewed class distributions \cite{jia2024, feng2024}. For example, in medical applications such as predicting outcomes in ICU patients, achieving high precision and recall is critical to ensure that significant cases are not missed, thereby improving patient care \cite{iwase2021}. These metrics are also widely used in other fields, such as healthcare and cybersecurity, where the consequences of misclassification can be severe \cite{abbas2023}. Therefore, understanding and employing these metrics is crucial for developing reliable ML models that can perform effectively in diverse applications.

\section{Experiment and results}
\label{experiment}
\begin{figure*}[h]
\centering
\includegraphics[width=11.5cm]{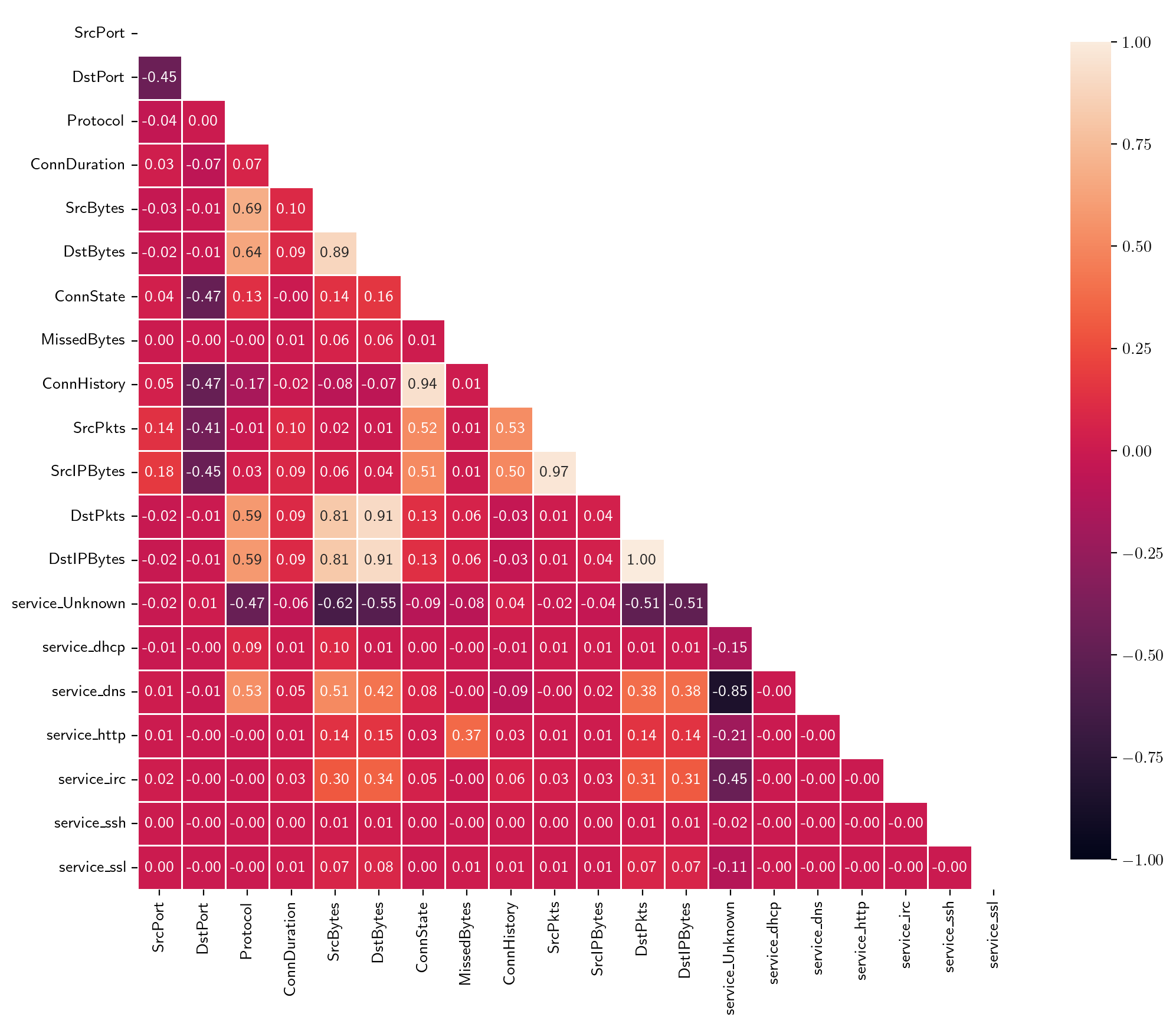}
\caption{Spearman's rank correlation of the dataset.}
\label{tab:spearman}
\end{figure*}

\subsection{Overview of Experimental Setup}

The experiment was designed to test the performance of the models in a typical Edge device used in IoT networks, using both clean and noisy data. A Windows 11 personal computer was used to preprocess the data required to train the models. The preprocessing step included handling missing values, removing duplicates, and addressing dataset inconsistencies. Feature selection was conducted using Spearman's rank correlation and ML-based methods, identifying key attributes for the models. Statistical analysis of feature distributions further informed the selection process.

Subsequent model experiments and evaluations were executed on a Raspberry Pi 5, leveraging its quad-core ARM Cortex processor (2.4 GHz) and 4 GB RAM. This setup allowed testing in resource-constrained environments, representative of IoT and edge-computing deployments. By combining preprocessing on a robust desktop system with model execution on the Raspberry Pi, the experiment ensured a comprehensive assessment of computational and performance trade-offs.

Each ML model was implemented using its respective framework or library. A deep learning model was also implemented using TensorFlow for comparison purposes and was trained and tested on the preprocessed data. Training was performed with hyperparameter tuning to optimize performance, followed by testing on a separate dataset to evaluate predictive accuracy and robustness. Model performance was assessed using precision, recall, and F1-score to provide a comprehensive comparison.

\subsection{Data Processing and Feature Engineering}

\subsubsection{Dataset Processing}

A balanced subset of 400,000 data samples from the IoT-23 dataset \cite{garcia2020iot23}, comprising 50\% malicious and 50\% benign instances, was selected for evaluation. Spearman's rank correlation analysis was conducted, with the correlation heatmap displayed in Figure~\ref{tab:spearman}.

\begin{figure*}
\centering
\includegraphics[width=13cm]{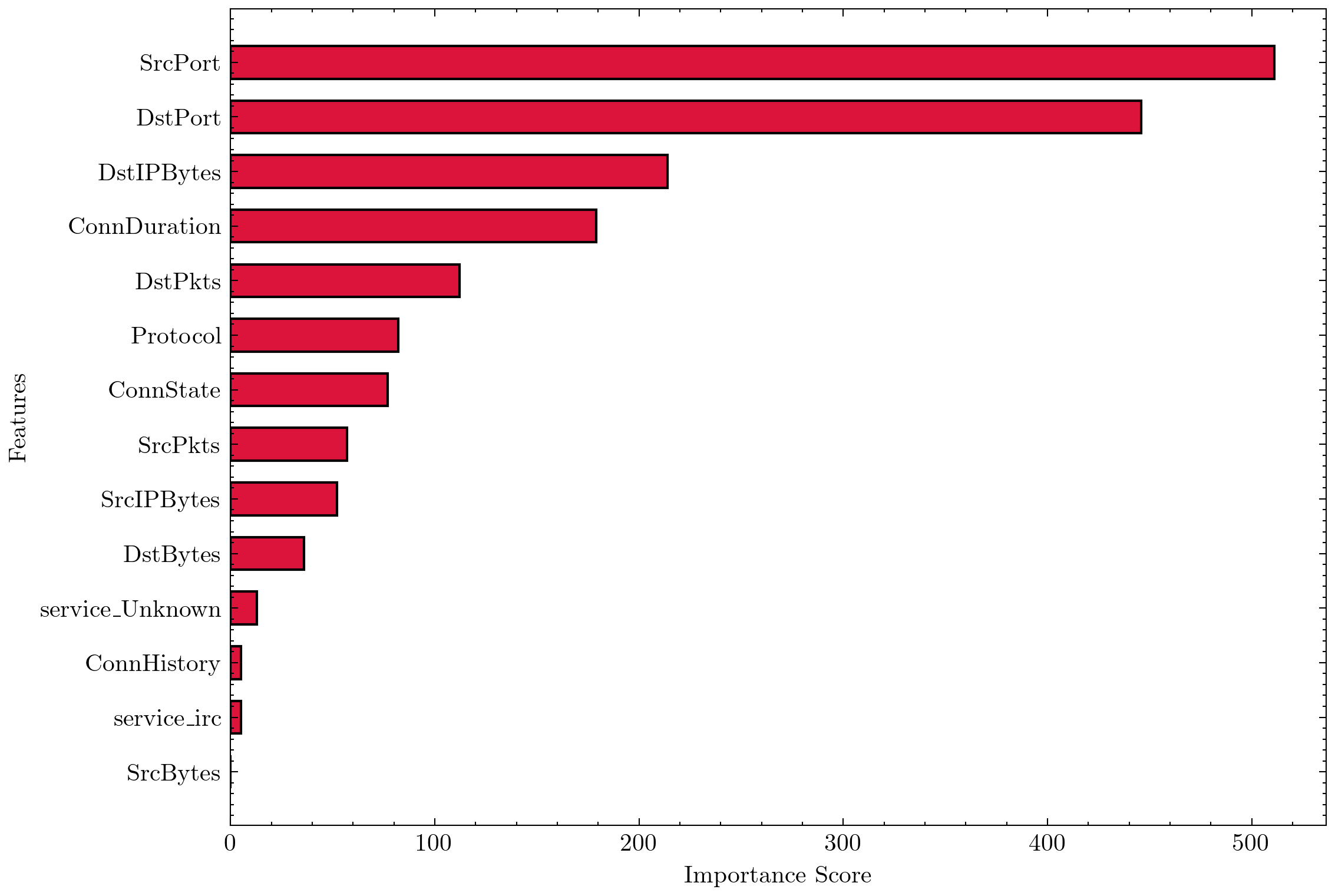}
\caption{Feature importance scores based on the XGBoost selection method.}
\label{fig:feature-importance}
\end{figure*}

The XGBoost feature importance analysis, as depicted in the Figure \ref{fig:feature-importance}, reveals that SrcPort (524.0) and DstPort (474.0) are the most influential features, indicating the critical role of source and destination port numbers in distinguishing between benign and malicious traffic. SrcIPBytes (209.0) and ConnDuration (175.0) further highlight the importance of data size and connection duration in detecting botnet activities. Additional features, such as SrcPkts (126.0), ConnHistory (84.0), and ConnState (49.0), provide valuable insights into traffic behavior and connection state. Features like Protocol (44.0), SrcBytes (39.0), and lesser-ranked ones, including service\_Unknown (16.0) and DstPkts (14.0), contribute to a lesser extent but still support the model's overall performance.

Feature importance analysis was performed using XGBoost, with results visualized in Figure~\ref{fig:feature-importance}. Features with non-zero importance scores were selected for further processing.

Categorical features were binary-encoded, and numerical features were scaled appropriately. The dataset was split into three subsets: 64\% for training, 16\% for validation, and 20\% for testing.

\subsubsection{Hyperparameter Tuning}

RandomizedSearchCV was employed for hyperparameter tuning, leveraging its ability to efficiently explore a predefined parameter space for the Random Forest, XGBoost, and LightGBM models. By sampling a fixed number of combinations from the parameter pool, RandomizedSearchCV balances computational efficiency with performance optimization. The selected hyperparameters for each model are detailed in Table~\ref{tab:hyperparameters}.

\begin{table}[h!]
    \centering
    \begin{tabular}{@{}lccc@{}}
        \toprule
        \textbf{Hyperparameter} & \textbf{Random Forest} & \textbf{XGBoost} & \textbf{LightGBM} \\
        \midrule
        max\_depth & 6 & 6 & 5 \\
        n\_estimators & 100 & 100 & 200 \\
        min\_samples\_split / min\_child\_samples & 10 & - & 20 \\
        min\_samples\_leaf & 4 & - & - \\
        max\_features & \textquotesingle sqrt\textquotesingle & - & - \\
        subsample & - & 1.0 & 0.9 \\
        learning\_rate & - & 0.1 & 0.01 \\
        num\_leaves & - & - & 31 \\
        reg\_alpha & - & - & 1.0 \\
        reg\_lambda & - & - & 0 \\
        colsample\_bytree & - & 0.8 & 0.9 \\
        gamma & - & 0.1 & - \\
        \bottomrule
    \end{tabular}
    \caption{Hyperparameters for the selected ML models.}
    \label{tab:hyperparameters}
\end{table}

\subsection{ML Model Evaluation}

This section examines the results and evaluates the performance of each ML model utilized in the experiment.

\subsubsection{Random Forest Classifier}

\begin{figure}[h!]
    \centering
    \subfloat[Validation Matrix]{\includegraphics[width=5cm]{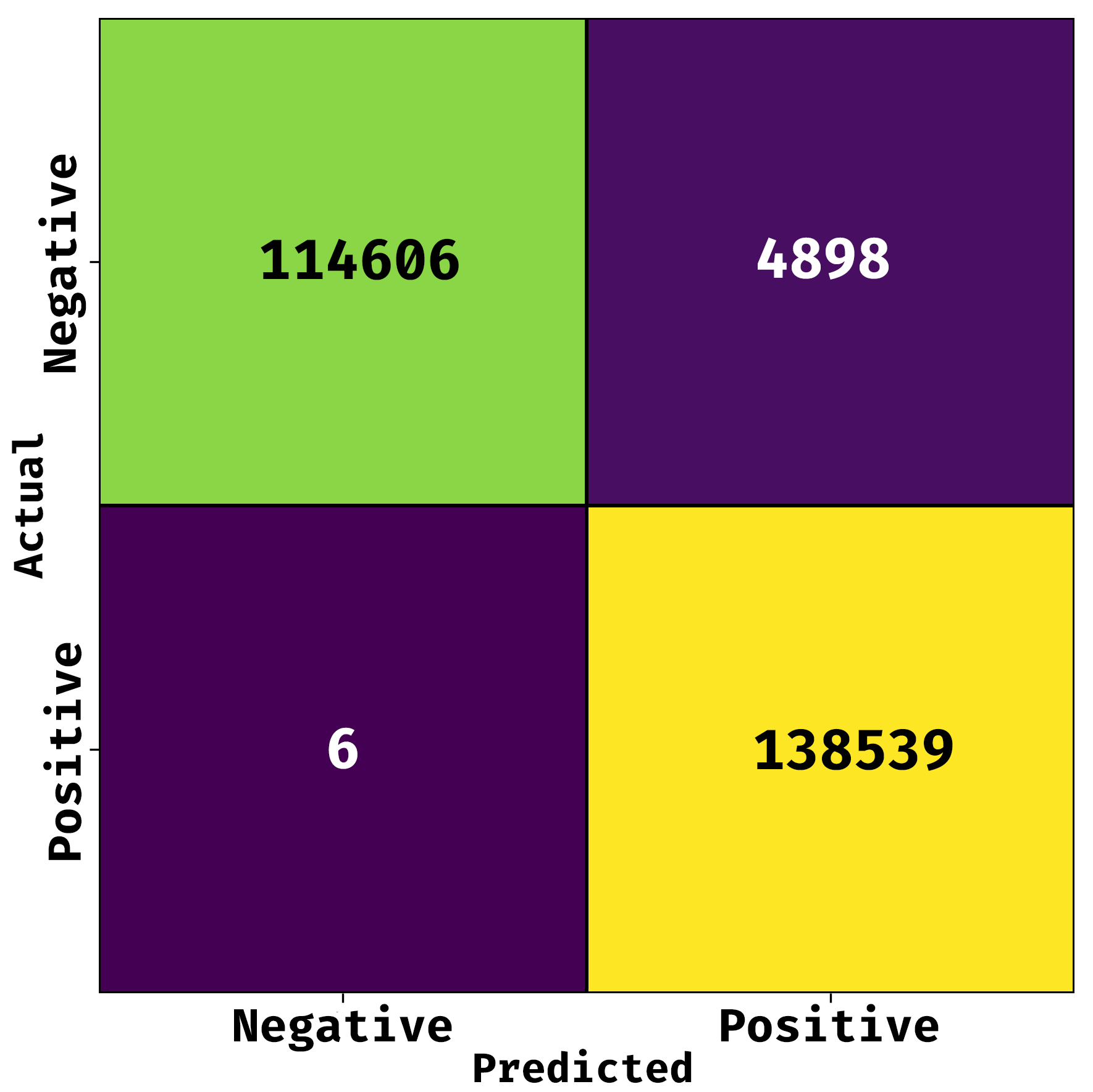}\label{fig:rf_val}}
    \qquad
    \subfloat[Test Matrix]{\includegraphics[width=5cm]{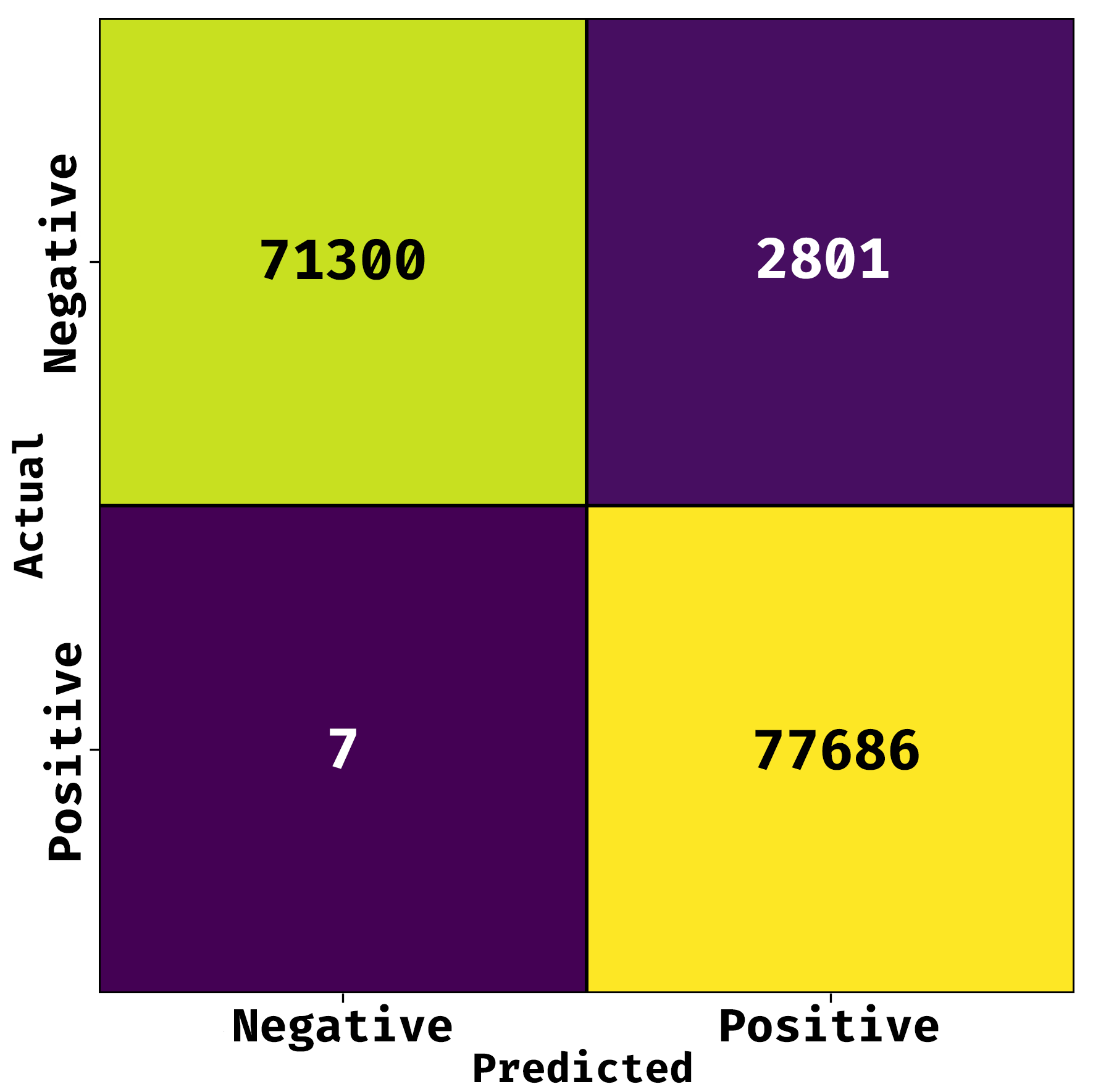}\label{fig:rf_test}}
    \caption{Random Forest classifier evaluation results.}
    \label{fig:rf_matrices}
\end{figure}

The confusion matrices in Figure \ref{fig:rf_matrices} for the Random Forest classifier show strong performance on the validation dataset. The model achieved an accuracy of 99.0\%, correctly classifying most samples with few errors. The precision of 96.6\% indicates that most positive predictions were correct, with 4.1\% being false positives. The recall of 99.996\% shows the model's ability to detect almost all actual positive cases with very few false negatives. The F1-score of 98.2\% reflects a good balance between precision and recall. The specificity of 95.9\% shows the model effectively identified negative cases with a low number of false positives. The false negative rate (FNR) was 0.0043\%, and the false positive rate (FPR) was 4.1\%. The model achieved similar accuracy on validation and test datasets and has a saved size of 2.8 MB.

\subsubsection{XGBoost Classifier}

\begin{figure}[h!]
    \centering
    \subfloat[Validation Matrix]{\includegraphics[width=5cm]{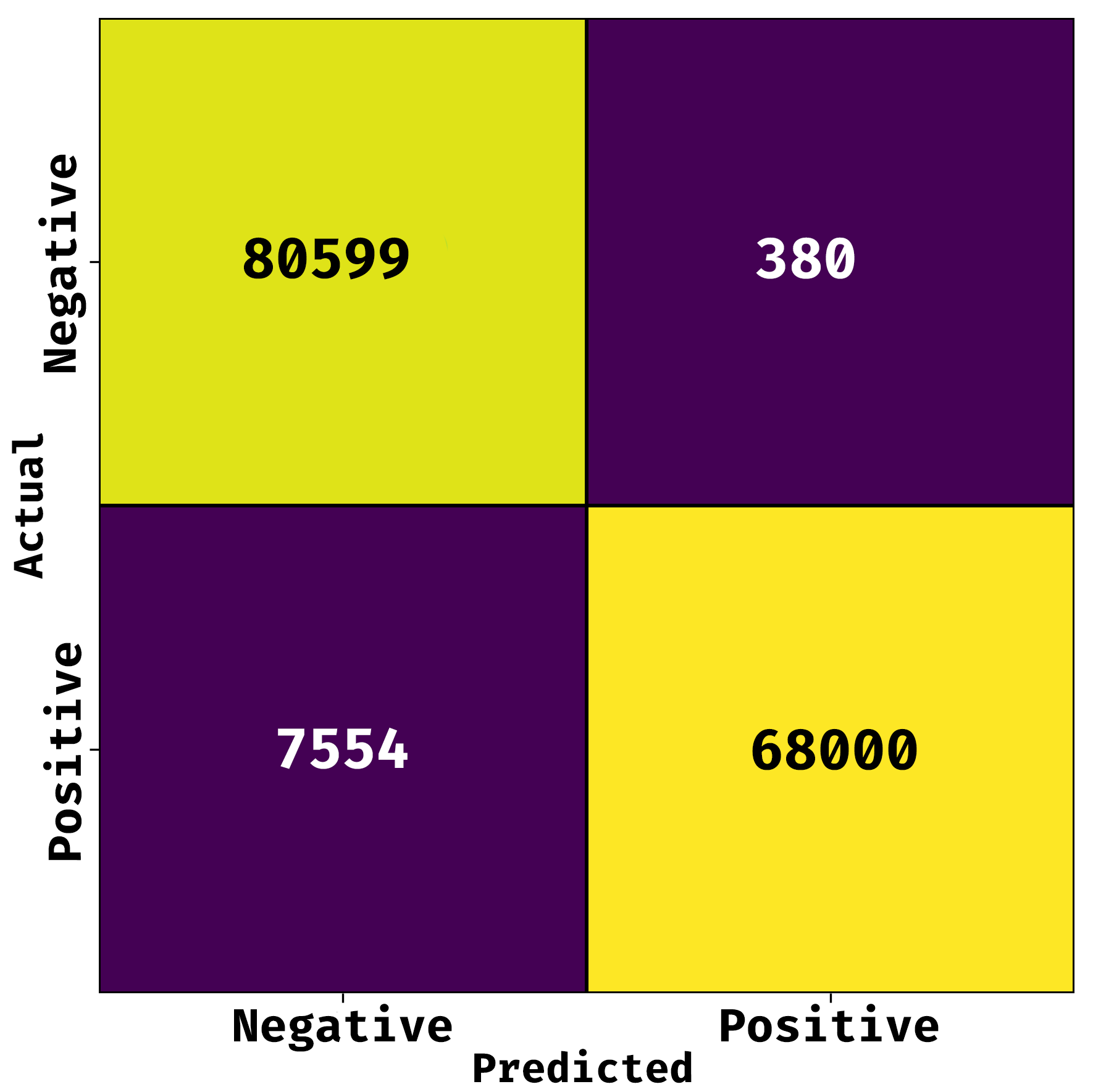}\label{fig:xgb_val}}
    \qquad
    \subfloat[Test Matrix]{\includegraphics[width=5cm]{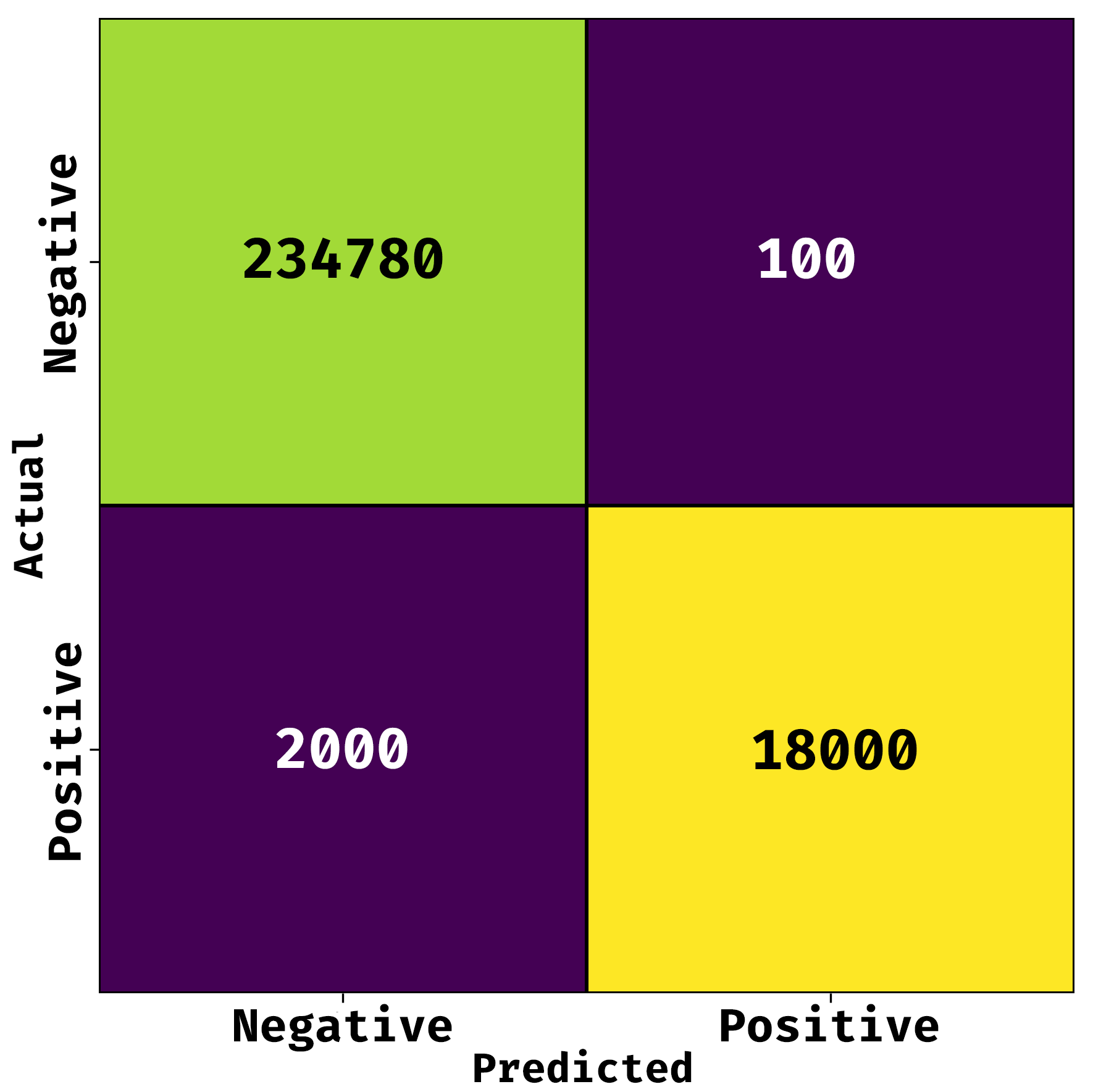}\label{fig:xgb_test}}
    \caption{XGBoost classifier evaluation results.}
    \label{fig:xgb_matrices}
\end{figure}

The confusion matrices in Figure \ref{fig:xgb_matrices} for the XGBoost classifier demonstrate its strong performance on both validation and test datasets. In the validation phase, the model achieved an accuracy of 97.9\%, with a precision of 99.45\% and a recall of 90\%, effectively identifying the majority of positive cases while maintaining a low number of false positives. The F1-score of 94.55\% reflects a solid balance between precision and recall, and the specificity of 99.87\% highlights its ability to classify negative cases accurately. On the test dataset, the model maintained similarly high performance, with metrics closely aligned to those observed in validation, further emphasizing its reliability. Additionally, the model's compact size of just 215 KB Figure~\ref{fig:xgb_matrices} showcases its efficiency alongside its strong performance.

\subsubsection{LightGBM Classifier}

\begin{figure}[h!]
    \centering
    \subfloat[Validation Matrix]{\includegraphics[width=5cm]{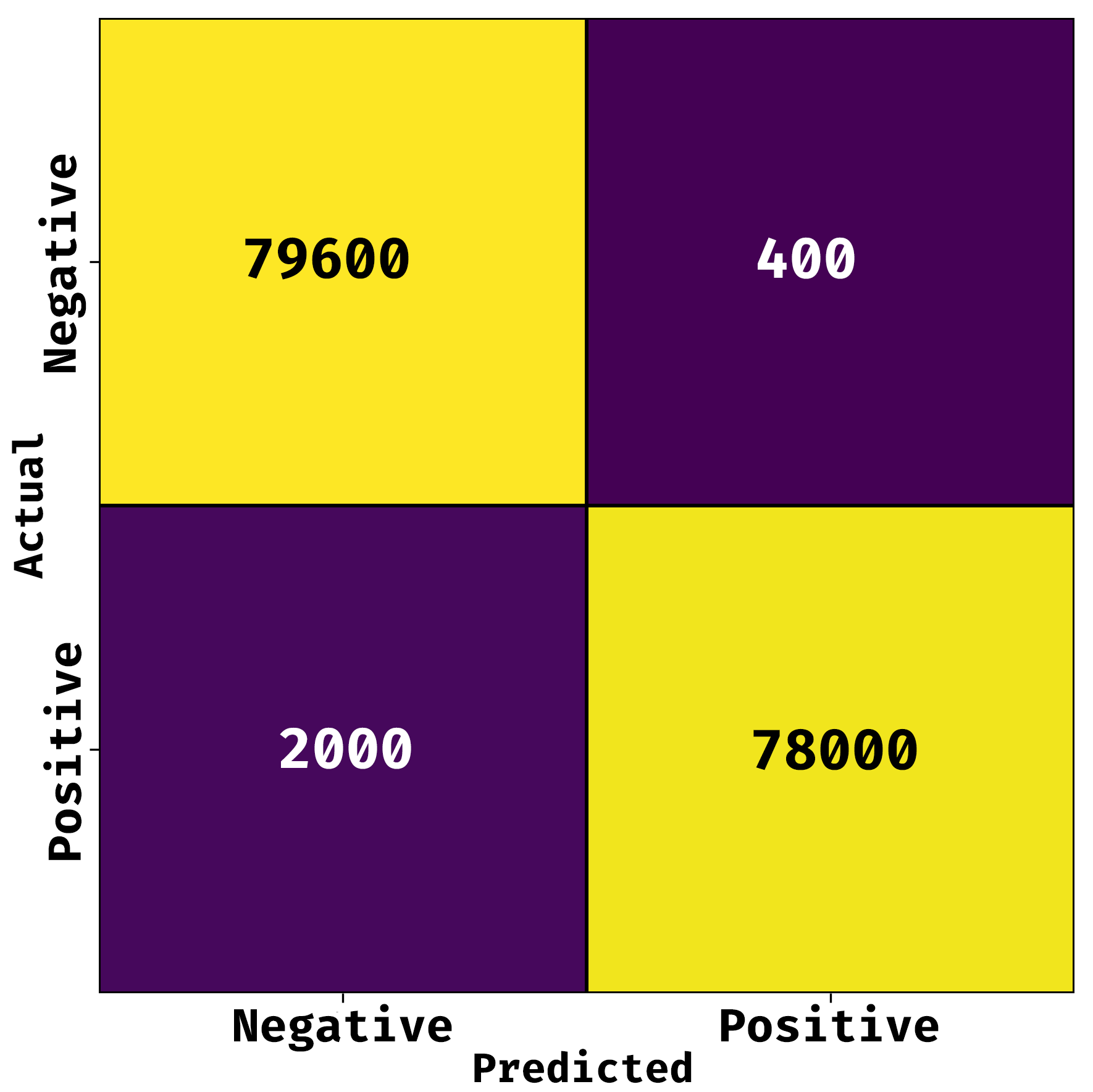}\label{fig:lightgbm_val}}
    \qquad
    \subfloat[Test Matrix]{\includegraphics[width=5cm]{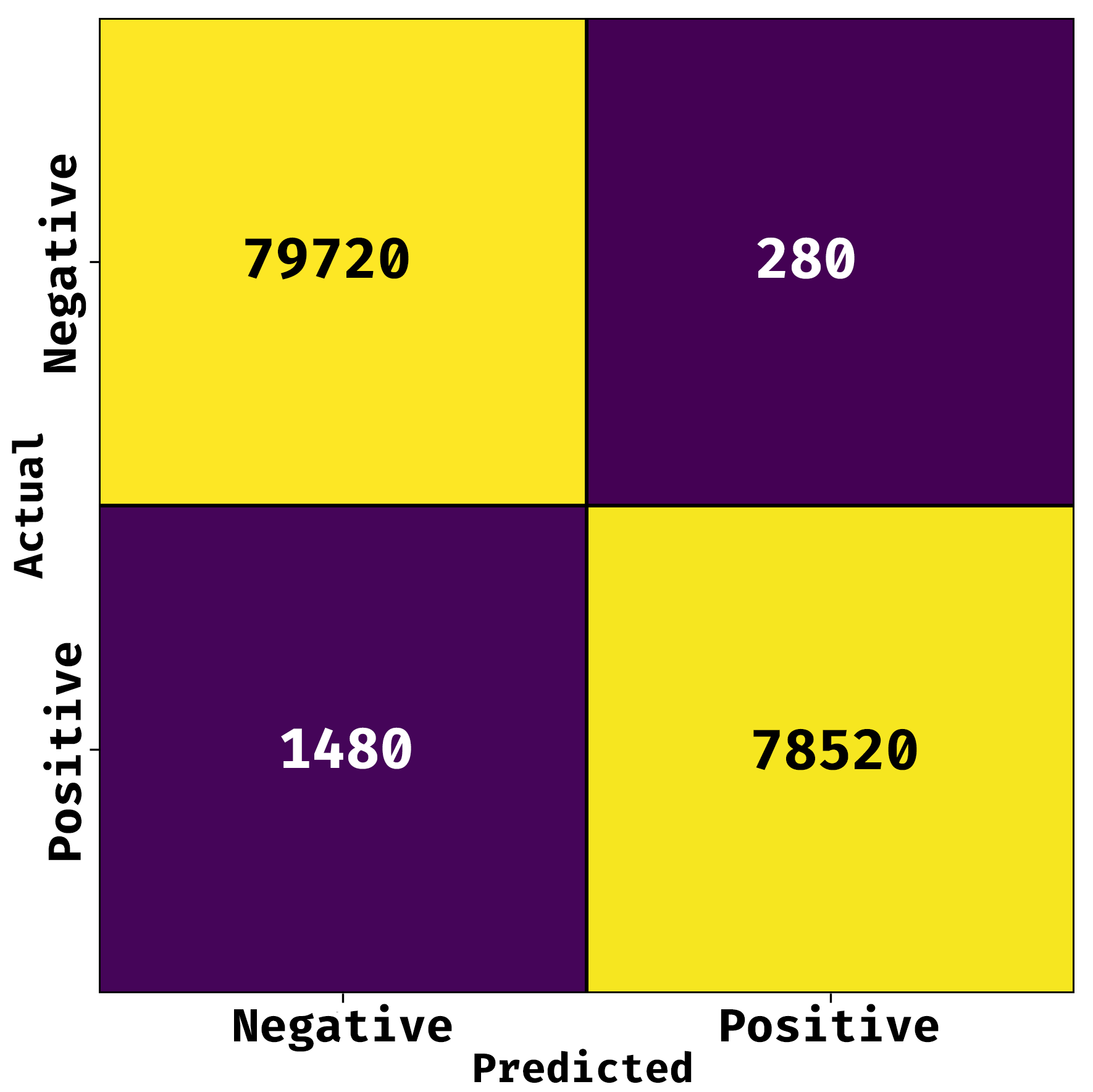}\label{fig:lightgbm_test}}
    \caption{Evaluation and test results of LightGBM classifier.}
    \label{fig:lightgbm_matrices}
\end{figure}

The confusion matrices for the LightGBM classifier, as shown in Figure \ref{fig:lightgbm_matrices}, demonstrate strong performance on both validation and test datasets. In the validation phase, the model achieved an accuracy of 98.7\%, with a precision of 99.5\% and a recall of 97.5\%, effectively identifying most positive cases while maintaining a low number of false positives. On the test dataset, the model delivered consistent results, with an accuracy of 98.7\%, precision of 99.6\%, and recall of 98.1\%. The F1-score of 98.5\% reflects a balanced performance between precision and recall. With a compact saved size of 541 KB, the LightGBM classifier is highly efficient and well-suited for resource-constrained environments, such as IoT and edge-computing scenarios.

\begin{figure*}
\centering
\includegraphics[width=12cm]{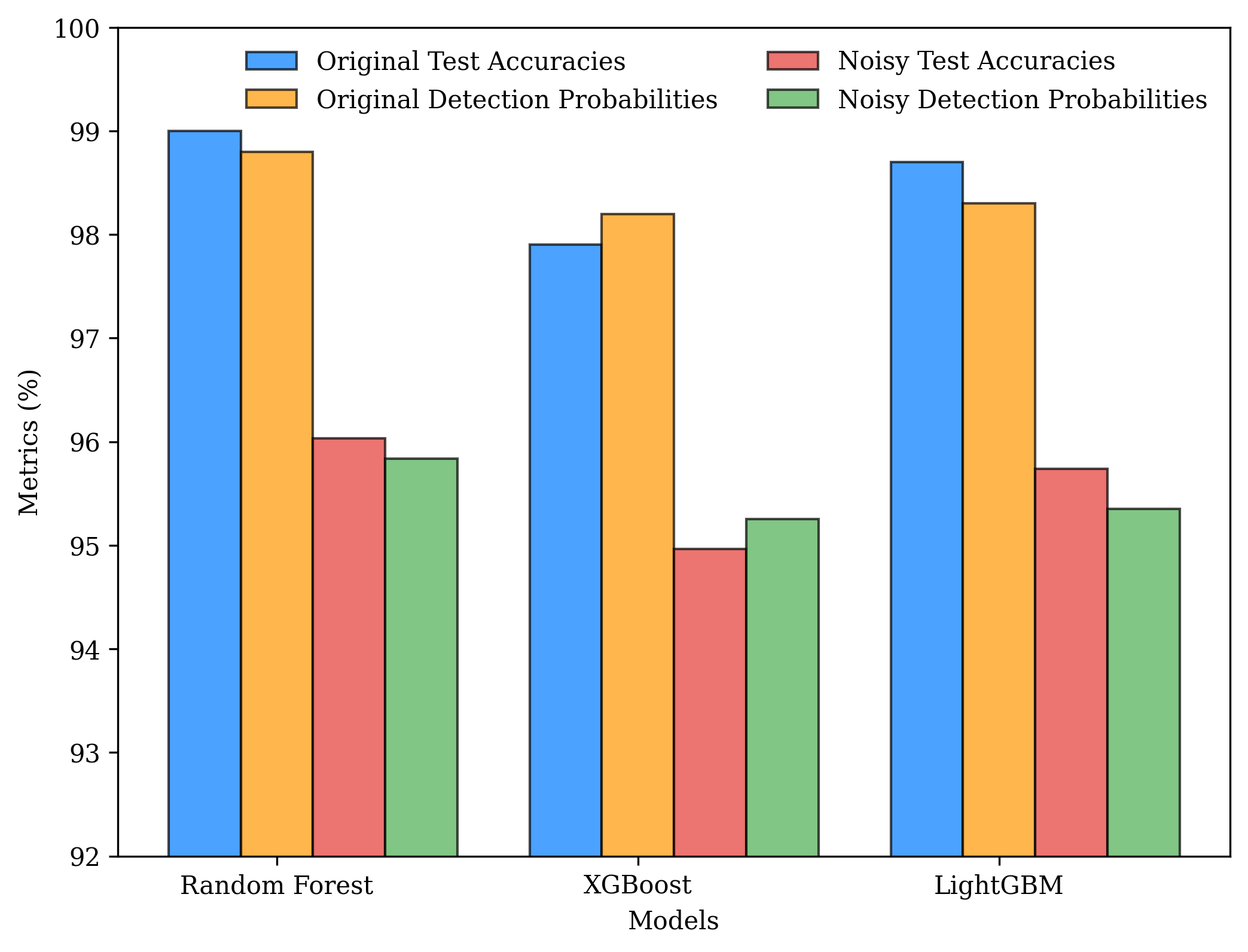}
\caption{Model accuracy and detection probability comparison across classifiers.}
\label{fig:model_accuracy}
\end{figure*}

\subsection{Performance metrics and discussion}

\begin{table}[h!]
    \centering
    \begin{tabular}{ccc}
        \toprule
        \textbf{Category} & \textbf{Test Accuracy (\%)} & \textbf{Detection Probability (\%)}\\
        \midrule
        Random Forest & 99.0 & 98.8 \\
        XGBoost & 97.9 & 98.2 \\
        LightGBM & 98.7 & 98.3 \\
        DFNN &  99.1 & 98.9 \\
        \bottomrule
    \end{tabular}
    \caption{Comparison of classifiers: test accuracies and detection probabilities.}
    \label{tab:classifier-comparison}
\end{table}

To assess real-world scenarios, Random Feature Corruption was applied to the test dataset, introducing noise. This included Gaussian noise, where small random values from a normal distribution were added to numerical features, simulating real-world data variations. The models were then re-evaluated on the modified dataset to measure their accuracy and detection probabilities, assessing their robustness against noisy and imperfect data.

As described above, a DFNN was designed with multiple fully connected layers to evaluate its feasibility against traditional ML models. The architecture consists of an input layer matching the feature dimensions and four hidden layers. Each hidden layer uses the ReLU activation function to introduce non-linearity, while batch normalization is applied to stabilize training. Dropout layers with rates of 0.4 and 0.3 help mitigate overfitting by randomly deactivating neurons during training. The output layer employs a softmax activation function, making the model suitable for multi-class classification. The DFNN was trained using categorical cross-entropy loss and optimized with the Adam optimizer to balance convergence speed and accuracy.

The chart in Figure \ref{fig:model_accuracy} compares the test accuracy and detection probability of three ML models using both preprocessed and noisy data. Overall, introducing noise reduces performance by about 3\% across all models. Table \ref{tab:classifier-comparison} provides a summary of these results. Among the ML models, Random Forest achieves the highest test accuracy and detection probability, both slightly above 99\%. LightGBM performs similarly in terms of accuracy but has a slightly lower detection probability. XGBoost records the lowest values in both metrics, indicating weaker performance in this scenario. The DFNN model outperforms the ML models, achieving a test accuracy of 99.1\% and a detection probability of 98.9\%.

Table~\ref{tab:model-performance} compares the performance of XGBoost, LightGBM, Random Forest, and DFNN in terms of training time, inference time, and model size using 15 features. LightGBM remains the most efficient choice for resource-constrained edge devices, with the shortest training time (0.79 s) and a relatively small model size (0.58 MB). XGBoost offers the fastest inference time (0.102 s) but has a significantly longer training time (6.62 s). Random Forest provides balanced performance but has the largest model size (2.8 MB), which may be a limitation for edge deployments. DFNN achieves high accuracy but comes with a substantial increase in training time (76.59 s) and a larger inference time (0.241 s), making it less suitable for real-time edge applications.

\begin{table}[]
    \centering
    \begin{tabular}{@{}lcccc@{}}
        \toprule
        \textbf{Model} & \textbf{Features} & \textbf{Training Time (s)} & \textbf{Inference Time (s)} & \textbf{Model Size (MB)} \\
        \midrule
        XGBoost        & 15                          & 6.627                       & 0.102                       & 0.21                     \\
        LightGBM       & 15                          & 0.791                     & 0.161                      & 0.58                     \\
        Random Forest  & 15                          & 3.568                     & 0.178                      & 2.80                     \\
        DFNN  & 15                          & 76.590                    & 0.241                     & 0.91                     \\
        \bottomrule
    \end{tabular}
    \caption{Model performance comparison including training time, inference time, and model sizes.}
    \label{tab:model-performance}
\end{table}

The LightGBM model emerges as a particularly effective choice for deployment in constrained environments, such as IoT and edge-computing scenarios. Its performance metrics, including a high test accuracy of 98.7\%, precision of 99.6\%, and recall of 98.1\%, demonstrate its reliability and robustness in detecting critical patterns within the data. Moreover, LightGBM's compact model size of just 541 KB makes it highly efficient in terms of memory usage compared to other classifiers like Random Forest. This efficiency, combined with its ability to maintain strong predictive performance on resource-limited hardware, underscores its suitability for real-world applications where computational and storage resources are at a premium. The balance of accuracy, speed, and lightweight architecture positions LightGBM as an optimal choice for modern, constrained environments.

\section{Conclusion}
\label{conclusion}
This study examined the use of ML techniques for detecting botnet attacks in edge-computing-assisted IoT networks. Using a subset of the IoT-23 dataset, four models—Random Forest, XGBoost, LightGBM, and a DFNN —were implemented and evaluated based on accuracy, computational efficiency, and resource usage in constrained environments. While all models performed well, LightGBM emerged as the most practical choice due to its high accuracy, minimal resource consumption, and scalability, making it particularly effective for real-time detection on edge devices. Additionally, the study examined the impact of noisy data, revealing a consistent performance drop of approximately 3\% across all models. Despite achieving the highest accuracy, DFNN required significantly more computational resources, making it less suitable for edge-device deployment. These findings highlight the trade-offs between deep learning and lightweight ML models in resource-limited environments. This work reinforces the potential of efficient machine learning models in enhancing IoT network security while balancing performance and resource constraints. Future research could further optimize these models, refine deep learning architectures for edge compatibility, and test their resilience against various real-world IoT attack scenarios, including adversarial noise.
\bibliographystyle{unsrt}  
\bibliography{references}

\end{document}